\documentclass[12pt]{iopart}
\usepackage[lofdepth,lotdepth]{subfig}
\usepackage{graphicx}
\usepackage{color}
\usepackage{ulem}
\usepackage{cite}
\usepackage{hyperref}
\usepackage{algpseudocode}
\usepackage{amssymb}
\usepackage{caption}
\usepackage{upgreek}
\usepackage{multirow}
\usepackage{diagbox}
\usepackage{slashbox}
\usepackage{booktabs}
\usepackage{hhline}

\newcommand{\vJ}{{\bf J}}
\newcommand{\vE}{{\bf E}}

\newcommand{\vA}{{\bf A}}

\newcommand{\half}{\frac{1}{2}}

\begin{document}

\title{Modeling the charging process of a coil by an HTS dynamo-type flux pump}
\markboth{Modeling the charging process of a coil by an HTS dynamo ...}{}

\author{$\textrm{Asef Ghabeli}^{1*},\:\textrm{Mark Ainslie}^2,\:  \textrm{Enric Pardo}^1,$ \\
$ \textrm{Lo\"ic Qu\'eval}^3 \: \textrm{and} \: \textrm{Ratu Mataira}^4$  
}

\address{1 Institute of Electrical Engineering, Slovak Academy of Sciences, Bratislava, Slovakia\\ 
2 Department of Engineering, University of Cambridge, United Kingdom \\
3 Group of Electrical Engineering Paris (GeePs), CentraleSup\'elec, University of Paris-Saclay, France \\
4 Robinson Research Institute, Victoria University of Wellington, New Zealand\\}

\ead{asef.ghabeli@savba.sk}

\date{\today}

\begin{abstract}

The high-$T_c$ superconducting (HTS) dynamo exploits the nonlinear resistivity of an HTS tape to generate a DC voltage when subjected to a varying magnetic field. This leads to the so-called flux pumping phenomenon and enables the injection of DC current into a superconducting coil connected to the dynamo without current leads. In this work, the process of charging a coil by an HTS dynamo is examined in detail using two numerical models: the Minimum Electromagnetic Entropy Production and the segregated $\textbf{H}$-formulation finite element model.~The numerical results are compared with an analytical method for various airgaps and frequencies.~Firstly, the \textit{I-V} curves of the modeled HTS dynamo are calculated to obtain the open-circuit voltage, short-circuit current and internal resistance.~Afterward, the process of charging a coil by the dynamo including the charging current curve and its dynamic behavior are investigated. The results obtained by the two models show excellent quantitative and qualitative agreement with each other and with the analytical method.~Although the general charging process of the coil can be obtained from the \textit{I-V} curve of the flux pump, the current ripples within a cycle of dynamo rotation, which can cause ripple AC loss in the HTS dynamo, can only be captured via the presented models.       

\end{abstract}

\maketitle

\section{Introduction}  

	High-$T_c$ superconducting (HTS) flux pumps employ a source of varying magnetic field relative to an HTS tape to generate a DC voltage (with a large ripple) and inject DC current into a superconducting coil connected to it. In HTS dynamo-type flux pump, this source of varying magnetic field is a permanent magnet(s) that rotates around an axis and transits past a stationary superconducting tape(s). The working mechanism of an HTS dynamo is similar to traditional dynamos, but without the need for commutator and brushes. Indeed, in the HTS dynamo, the nonlinear resistivity of superconductor serves as a natural rectifier to generate a DC voltage within each cycle. 

	The HTS dynamo has drawn much attention in the last decade, since it was proposed and designed by Hoffman \textit{et al.} in 2011 \cite{hoffmann2010flux}. It has a simple structure and low maintenance compared to other types of flux pumps. They can inject DC current into superconducting magnets \cite{HoffmannC2012PP,walshRM2014IES} or the rotor winding of electrical machines without the need for brushes or bulky current leads and their associated thermal loss. This will reduce the maintenance of electrical machines and increase the efficiency of cryogenic systems\cite{haran2017high,bumby2016development,bumby2016through,JiangZ2016IES,badcock2016impact,pantoja2016impact,StoreyJ2018IES,hamilton2018design, storey2019optimizing,coombs2019superconducting, HamiltonK2020IES}. Such flux pumps can be also useful to charge no-insulation coils \cite{MaJ2019IES,GengJ2019SUST}.  

	Many articles have reported experimental investigations into the process of charging a superconducting coil using an HTS dynamo, starting with \cite{hoffmann2010flux}. In \cite{ZhenanJ2014APL}, it was shown that the maximum output current is limited by the dynamic resistance, which can be minimized by optimizing key design parameters of the flux pump. In \cite{ZhenanJ2015SUST}, the impact of airgap and in \cite{JiangZ2016IES,bumby2016through,StoreyJ2018IES}, the impact of using a ferromagnetic circuit with varying yoke width, airgap and frequency on the charging performance of an HTS dynamo were studied. In \cite{pantoja2016impact}, the impact of stator wire width and in \cite{HanS2019IES} impact of HTS wire type and frequency on charging of HTS coil were investigated. The dynamic charging current curve of a flux pump is not smooth as it contains ripples within each cycle. The source of these ripples were analyzed and discussed by experiments in a transformer-rectifier HTS flux pump in \cite{GengJ2019SUSTAKilo} and in a pulse-type magnetic flux pump in \cite{BaiZ2017IES}. However, there is no experimental work regarding the source of the current ripple in an HTS dynamo.  

Although several experimental studies have been carried out to explore the impact of various design parameters of an HTS dynamo on the charging process of an HTS coil, there is still a need for efficient models to fully examine the details and optimize this process. In recent years, several numerical models have been developed, which can be categorized into two groups. The first group models the open-circuit mode, where the output current is zero \cite{GhabeliA2020SUST, mataira2019origin, ainslie2020new, ghabeli2020arXiv, prigozhinL2020SUST, AinslieM2021IES,prigozhinL2020IES}. All of these models could explore the essential mechanism of the flux pump to deliver a DC voltage by the assumption of constant critical current density $J_c$ for the HTS tape characteristic. However, only some of them considered $J_c(B,\theta)$ dependency, which enable the models to generate an output voltage much closer to reality and comparable to experiments \cite{GhabeliA2020SUST, mataira2019origin,  ghabeli2020arXiv}. The second group presented in \cite{matairaR2020PRA} and \cite{matairaR2020IES} are capable of also modeling the HTS dynamo with an imposed DC transport current; using these models, the \textit{I-V} curves of the flux pump and the associated effective resistance at different frequencies could be obtained. 

Proper modeling of the dynamic behavior is important because the ripples within each cycle can potentially cause AC loss in the coil. In addition, some applications such as motors and generators can be sensitive to ripple currents and their ripple magnetic fields.          

In this work, we model the charging process of a coil by an HTS dynamo-type flux pump. We perform these calculations using both the minimum electromagnetic entropy production (MEMEP) method and the segregated $\textbf{H}$-formulation finite element method, benchmarking the two techniques. The numerical results are also compared to analytical results. The models allows us to study the current ripples and their resultant AC loss in an HTS dynamo during the charging process.    

\section{Problem Configuration}
\begin{figure}[tbp]
\centering
{\includegraphics[trim=0 0 0 0,clip,width=7.5 cm]{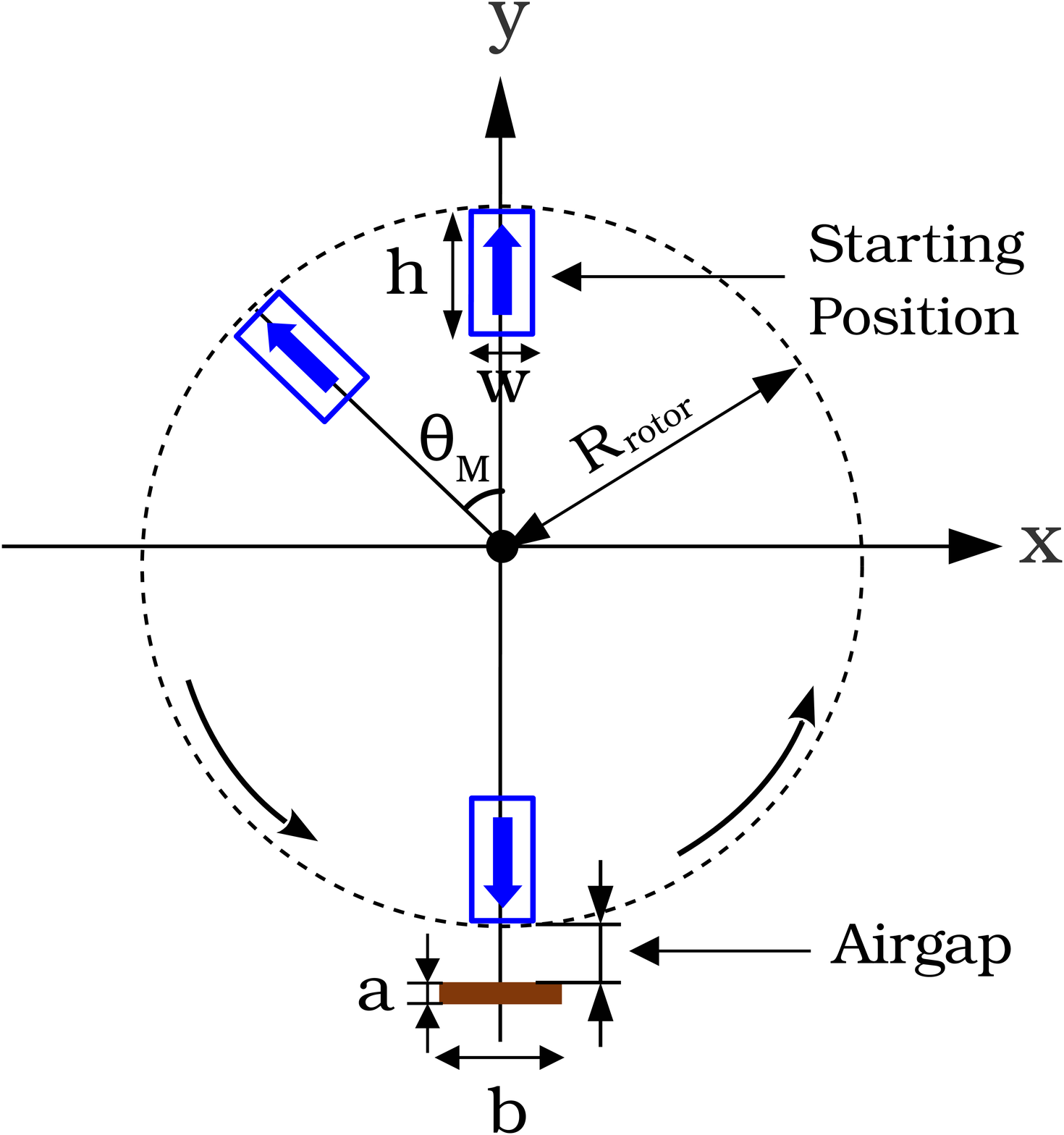}}
\caption {Configuration of the studied problem in the $xy$-plane .} 
\label{fig.configuration}
\end{figure}

Fig. \ref{fig.configuration} shows the configuration of the studied problem.\;The permanent magnet with width $w$, height $h$ and remanent flux density $B_r$ rotates in the $xy$-plane in the counterclockwise direction with its magnetization facing outwards. $\theta_M$ is the magnet angle of rotation and the airgap is defined as the minimum distance between the magnet outer surface and the tape surface at $\theta_M=180^{\circ}$. The HTS tape has width $b$, thickness $a$ and critical current density $J_c$, and the temperature assumed to be constant. The tape effective length (depth) $l$ is defined as the length of the tape and magnet in the $z$ direction, which is used to calculate the voltage, as per Equation (\ref{V_instan}).

In our study, a constant $J_c$ equal to $J_{c0}=I_{c0}/(b\cdot a)$ has been assumed for simplicity. Although this will reduce the charging efficiency of the flux pump as discussed in \cite{CampbellA2019SUST, matairaR2020PRA, GhabeliA2020SUST,AinslieM2021SUSTCorri}, it does not have any effect on the essential behavior of the flux pump to deliver a DC voltage.\;In addition, for simplicity and increasing the calculation speed, only the superconducting layer of the HTS tape was considered, which has a negligible effect on the performance of the flux pump at low frequencies \cite{BumbyC2017IES, PantojaA2018IES,AinslieM2021IES}. In our calculations, we assume an ideal HTS coil without considering its critical current or its dynamic effects due to screening currents. It is modeled as lumped parameter elements including the inductance $ L=0.24 $\,mH and the joint resistance $R_c=0.88$\,$\mu\Omega$. These values were derived based on the experimental values presented in \cite{ZhenanJ2014APL}. However, for faster charging of the coil, the inductance $L$ was chosen as $1/10\textsuperscript{th}$ of its real value. The characteristics of the permanent magnet, the HTS tape and the HTS coil are summarized in Table \ref{table_parameters}, which are based on the HTS dynamo benchmark problem presented in \cite{ainslie2020new} and are derived from the experimental study performed in \cite{badcock2016impact}.

\begin{table}
\caption{Problem configuration parameters}
\centering
\label{table_parameters}
\resizebox{9 cm}{!}{
\begin{tabular}{ccc}
\hline
\hline
\multirow{4}{*}{Permanent Magnet} & Width, $w$                  & 6 mm \\
                                  & Height, $h$                 & 12 mm \\
                                  & Effective length (Depth), $l$         & 12.7 mm \\
                                  & Remanent flux density, $B_r$ & 1.25 T \\ 
\hline
\multirow{4}{*}{HTS Tape}         & Width, $b$                  & 12 mm \\
                                  & Thickness, $a$              & 1 $\mu$m \\
                                  & Critical current $I_c$       & 283 A \\
                                  & n-value                     & 20 \\
\hline
\multirow{2}{*}{Charging Circuit} & Inductance, $L$                  & 0.24\,mH \\
                                  & Joint Resistance, $R_c$          & 0.88 $\mu\Omega$ \\ 
\hline
Rotor external radius, $R_{rotor}$ &                      & 35 mm     \\ 
Airgap                             &                      & 1, 2, 3.7 mm\\                        
Rotation frequency, $f$         &                      & 4.25, 25, 50 Hz \\

\hline
\hline
                                  
\end{tabular}
}
\end{table}

\section{Calculation Methods}
\subsection{General Definitions}
\label{General Definitions}
We assume an isotropic \textit{E-J} power law to define the non-linear characteristic of the HTS tape:

\begin{equation}
\textbf{E}(\textbf{J})=E_{c}\left(\frac{|\textbf{J}|}{J_{c}}\right)^n \frac{\textbf{J}}{|\textbf{J}|},
\label{E-J}
\end{equation}                                 
where $E_{c} =10^{-4}$ V/m is the critical electric field, $J_{c}$ is the critical current density and $n$ is the n-value, which defines the steepness of the transition between the superconducting state and the normal state. In this article, we assume that both tape and magnet are infinitely long along the $z$ axis in a Cartesian 2D coordinate system. Therefore, the current density $\vJ$, vector potential $\vA$, and electric field $\vE$, satisfy ${\textbf{J}}({\textbf{r}_2})=J(x,y)\,{\textbf{e}}_z$, ${\textbf{A}}({\textbf{r}_2})=A(x,y)\,{\textbf{e}}_z$ and ${\textbf{E}}({\textbf{r}_2})=E(x,y)\,{\textbf{e}}_z$, where $\textbf{r}_2$ is the 2D position vector in the cross-section and ${\textbf{e}}_z$ is the unit vector along the $z$ axis, corresponding to the assumed infinitely long direction of the tape. 

The relation between the current density and the vector and scalar potentials is: 
\begin{equation}
\textbf{E}(\textbf{J})=-\frac{\partial{\textbf{A}}}{\partial t}-\nabla\varphi, 
\label{E-phi}
\end{equation} 
and the current conservation equation: 
\begin{equation}
\nabla\cdot\textbf{J}=0, 
\label{nablaJ}
\end{equation}
where $ \partial {\textbf{A}}/\partial t$ is the change of vector potential with respect to time and $\varphi$ is the scalar potential. For Coulomb's gauge ($\nabla\cdot\textbf{A}=0$), the vector potential $\textbf{A}$ in Equation (\ref{E-phi})
includes the contributions from the applied field $\textbf{A}_{M}$ and the current density in the superconductor $\textbf{A}_{J}$. In addition, $\varphi$ in this gauge becomes the electrostatic potential (see the appendix of \cite{Grilli2014computation}). 

The cross-sectional average of the electric field $\textbf{E}(\textbf{J})$ is:
\begin{equation}
E_{av}(t)= \frac{1}{S} \int_{S_S} d^2 \textbf{r}_2\, \rho \big[J(\textbf{r}_2)\big] J(\textbf{r}_2),  
\label{Eq.Eav}
\end{equation}
where $S_s$ is the cross-section of superconducting tape, $d^2 \textbf{r}_2$ is the surface differential in the cross-section and $\rho$ is the nonlinear resistivity of the HTS tape, which can be obtained from Equation (\ref{E-J}). 

The total output voltage of the flux pump $V(t)$ is comprised of three components:
\begin{equation}
V(t)=-l\cdot\partial_{z}\varphi=l\cdot[E_{av}(J)+\partial_{t}A_{av}]=l \cdot \left[E_{av}(J)+\partial_{t}A_{M,av}+\partial_{t}A_{J,av}\right],
\label{V_instan}
\end{equation}                                
where $\varphi$ is the electrostatic potential, $E_{av}$ is obtained by Equation (\ref{Eq.Eav}), $A_{M,av}$ and $A_{J,av}$ are the average magnetic vector potential over the tape cross-section due to the permanent magnet and the screening current in the superconducting tape, respectively. Since $\partial_{z}\varphi$ is uniform within the superconductor, $E(J)$ and $\partial_{t}A$ can be written as the average over the tape cross-section, $E_{av}(J)$ and $\partial_{t}A_{av}$, respectively \cite{GhabeliA2020SUST}.  

Among these three terms, only $E_{av}$ is not periodic within a cycle leading to a DC voltage value and thus the pumping phenomenon. The two other terms $A_{M,av}$ and $A_{J,av}$ are periodic within a cycle and do not have any effect on the DC voltage value \cite{GhabeliA2020SUST,mataira2019origin}. However, they cause a significant ripple in the voltage and current, resulting in AC loss in the superconducting coil.

The cumulative total output voltage, $V_{cumul}(t)$, is:
 \begin{equation}
V_{cumul}(t)= \int_{0}^{t} V(t')\,dt'.
\label{V_cumul}  
\end{equation}
  
The DC output voltage of the dynamo, $V_{DC}$, is:
\begin{equation}
V_{DC}=\frac{1}{T} \int_{t}^{t+T} V(t')\,dt',  
\end{equation}
as the time average value of the induced voltage over one period of rotation, $T$, in the steady-state. 

\subsection{MEMEP 2D method}  \label{MEMEP method}

The MEMEP 2D method is a variational method, which the solution minimizes the entropy production made by electromagnetic fields {\cite{pardoE2015SUST,pardoE2017JCP}}.~It works based on the calculation of the current density $ \textbf{J} $, which only {exists} inside the superconducting {(or normal conducting)} region, thus, the discretization of mesh is only needed inside this region. For solving the problem, a functional needs to be minimized containing all the variables of the problem such as the magnetic vector potential $\textbf{A}$, current density $\textbf{J}$, and scalar potential $\varphi$. {In this article, we extend the numerical method in \cite{pardoE2015SUST,pardoE2017JCP} in order to take lumped circuit elements such as inductances and resistances into account.}

As detailed in \cite{pardoE2017JCP}, solving Eequation (\ref{E-phi}) in Coulomb's gauge is the same as minimizing the following functional:
\begin{eqnarray}
F & = & \int_{\Omega} d^3 \textbf{r} \bigg[ \half\frac{\Delta\textbf{A}_J}{\Delta t}\cdot\Delta\textbf{J}+\frac{\Delta\textbf{A}_M}{\Delta t}\cdot\Delta\textbf{J}+U(\textbf{J}_0+\Delta\textbf{J}) \nonumber \\
&& +\nabla\varphi\cdot(\textbf{J}_0+\Delta\textbf{J}) \bigg] ,
\label{F}
\end{eqnarray}
where $\Omega$ is the superconducting or normal conducting region in the 3D space, $\Delta\textbf{J}$, $\Delta\textbf{A}_J$, $\Delta\textbf{A}_M$ are the change of the variable between two consecutive time steps, $\Delta t$ is the time difference between two time steps, and $\textbf{J}_0$ is the current density at the previous time step. In addition, $U$ in this functional is the dissipation factor, defined as \cite{pardoE2017JCP}
\begin{equation}
U(\textbf{J})=\int_{0}^{\textbf{J}}\textbf{E}(\textbf{J}')\cdot d\textbf{J}'.
\label{dissipation}
\end{equation} 
This dissipation factor can include any \textit{E-J} relation for superconductors or normal conductors.

When the superconducting tape of the dynamo is connected in series with a coil, we can still use the functional of Equation (\ref{F}). In that case, $\Omega$ includes all parts: the superconducting tape, the coil and any series connected resistance (joints, coil resistance and so on). If the coil, resistance and tape are far away from each other, the magnetic field and vector potential from each component does not influence the other. Then  
\begin{equation}
F=F_S+F_L+F_R ,
\end{equation}
where $F_S$, $F_L$, $F_R$ are the terms (of a single functional) for the superconducting tape, coil, and resistance, respectively. The expression of these is the same as in Equation (\ref{F}) but replacing $\Omega$ by $\Omega_S$, $\Omega_L$ and $\Omega_R$, being the 3D regions for each part.

Next, we take several assumptions to simplify the 3D formulation of these functionals.

With the assumptions of section \ref{General Definitions}, $\textbf{J}(\textbf{r}_2)=J(x,y)\,\textbf{e}_z$, $\textbf{A}(\textbf{r}_2)=A(x,y)\,\textbf{e}_z$ and $\textbf{E}(\textbf{r}_2)=E(x,y)\,\textbf{e}_z$, where $\textbf{r}_2$ is the 2D position vector in the cross-section, $\textbf{r}_2=x\textbf{e}_x+y\textbf{e}_y$. In addition, $\nabla\varphi(\textbf{r})$ is uniform with value $\nabla\varphi(\textbf{r})=-(V/l)\textbf{e}_z$, where $\textbf{e}_z$ is the unit vector along the $z$ axis and $l$ is the tape effective length. The voltage $V$ is conventionally defined as in a passive circuit element $V=\varphi(z=-l/2)-\varphi(z=+l/2)$. With these assumptions, we simplify the superconducting tape functional into
\begin{eqnarray}
F_S & = & l\int_{S_S} d^2 \textbf{r}_2 \bigg [ \half\Delta J\frac{\Delta A_J}{\Delta t} + \Delta J\frac{\Delta A_M}{\Delta t}+U(J) \bigg] - V_SI,
\label{FS}
\end{eqnarray}
where $S_S$ is the cross-section of the superconducting tape and $I$ is the net current in the tape. The non-uniform applied magnetic field $\textbf{B}_{M}$ caused by the rotating magnet appears in the functional in the form of $A_{M}$. As well, in the infinitely long geometry, the $A_J$ contribution of the vector potential in Coulomb's gauge is \cite{Grilli2014computation}
\begin{equation}
A_J(\textbf{r}_2)=\frac{\mu_0}{2\pi}\int_{S_S} d^2 \textbf{r}_2'J(\textbf{r}_2)\ln|\textbf{r}_2-\textbf{r}_2'|,
\label{AJ}
\end{equation}

Although we may consider a superconducting coil with local non-linear resistivity, here we assume for simplicity an ideal inductor of inductance $L$. Thus, we neglect (non-linear) eddy currents in the conductor and any linear or non-linear resistive effect. As a result, $\textbf{J}$ is uniform in the conductor cross-section and $U(\textbf{J})=0$. Taking also into account that the coil is not submitted to external magnetic fields, the coil functional from the Equation (\ref{F}) reduces to
\begin{equation}
F_L=\half L \frac{(\Delta I)^2}{\Delta t}-V_LI , 
\label{FL}
\end{equation}
where the voltage of the inductance $V_L$, is conventionally defined as the difference in electrostatic potential between the entry and exit of current $I$ at the coil terminal.

In a similar way, we assume that there are no eddy currents in the resistance (representing joints), the resistivity there is linear, and that they are not submitted to external applied magnetic fields. Following the reasoning of \cite{liS2020SST}, we find that the resistance functional from the Equation (\ref{F}) reduces to
\begin{equation}
F_R=\half RI^2-V_RI
\end{equation}
for any 3D shape of the resistive joint, where $R$ is the resistance and the voltage $V_R$ is, again the drop in the electrostatic potential as defined in a passive circuit element. 

Since all elements are connected in series, the sum of the voltage terms of all three functionals is $I(V_S+V_L+V_R)$. Since the sum of the voltage drops are in a closed circuit, they follow $V_S+V_L+V_R=0$; being  Kirchhoff's second law. Thus, the whole system minimizes the following functional
\begin{eqnarray}
F & = & l\int_{S_S}d^2 \textbf{r}_2 \bigg [ \half\Delta J\frac{\Delta A_J}{\Delta t} + \Delta J\frac{\Delta A_M}{\Delta t}+U(J) \bigg] \nonumber \\
&& + \half L \frac{(\Delta I)^2}{\Delta t} + \half RI^2 .
\label{Fall}
\end{eqnarray}
Here, we have used the same voltage definition for all elements, as a passive circuit element, for consistency. However, if we consider the HTS dynamo as a voltage source, the voltage is defined with opposite sign. The latter definition is used in the analysis of section \ref{s.results}.

\subsection{Segregated $\textbf{H}$-formulation Finite Element Method}  \label{SEG-H 2D method}

The segregated $\textbf{H}$-formulation (SEG-H) finite-element model, implemented in COMSOL Multiphysics, consists of a magnetostatic permanent magnet model and a time-dependent \textbf{H}-formulation HTS wire model. The former is coupled unidirectionally to the latter using electromagnetic boundary conditions and a rotation operator to mimic the movement of the magnet \cite{queval2018superconducting,YangW2020IES,LiuK2017SUST}. This avoids the need for modeling moving parts (e.g., using a moving mesh) and significantly reduces the number of mesh elements, resulting in a fast and efficent model \cite{KajikawaK2003IES}.

The HTS wire model implements the 2D \textbf{H}-formulation \cite{KajikawaK2003IES,ainslie2020new,HongZ2006SUST,BrambillaR2006SUST,ainslieM2011comparison,AinslieM2012PhC}, where the independent variables are the components of the magnetic field strength $\textbf{H}$, and the governing equations are derived from Ampere's and Faraday's laws. On the outer boundary of the \textbf{H}-formulation model, the sum of the applied field $\textbf{H}_{ext}$, and the self-field $\textbf{H}_{self}$, is applied as a Dirichlet boundary condition. $\textbf{H}_{ext}$ is obtained by rotating the field of a static permanent magnet \cite{ainslie2020new} and $\textbf{H}_{self}$, created by the supercurrent flowing in the HTS wire, is obtained at each time step by numerical integration of the 2D Biot-Savart law over the HTS wire subdomain \cite{queval2018superconducting,LiuK2017SUST}.

The contribution, $A_{J}$, the vector potential due to the superconducting current, is calculated using COMSOL's Poission's Equation interface (one of the 'Classical PDEs' available in the Mathematics module), where
\begin{eqnarray}
\label{Eave}
\nabla\cdot(-\nabla A_{J})=\mu_0 J
\end{eqnarray}
and an appropriate Dirichlet boundary condition is set, on the outer boundaries of the HTS wire model, such that Equation (\ref{AJ}) is satisfied. 

The contribution, $A_{M}$ (see Equation (\ref{V_instan})\,), the vector potential due to the permanent magnet, is calculated using $A_z$ from the magnetostatic magnet model with the same rotation operator applied, as described earlier for the magnetic field, to mimic the movement of the magnet.

The total output voltage derived from the HTS wire model, including the contributions from $E_{av}$, $\partial_t A_{J,av}$ and $\partial_t A_{M,av}$, as defined by Equation (\ref{V_instan}), is then coupled to COMSOL's Electrical Circuit interface (AC/DC module) consisting of the voltage source (implemented using COMSOL's 'External I vs.V' node) in series with the inductance $L$ and resistance $R$. An ammeter ('Amp\'ere Meter' node) is also connected in series and the current flowing through this, $I_{cir}$, is coupled unidirectionally back to the HTS wire model with a constraint, such that

\begin{eqnarray}
\label{Eave}
I(t)=\int_{S}{J(t)\cdot dS}=I_{cir},
\end{eqnarray}

\subsection{Analytical Method}  \label{Analytical method}

\begin{figure}[tbp]
\centering
{\includegraphics[trim=0 0 0 0,clip, width=4.5 cm]{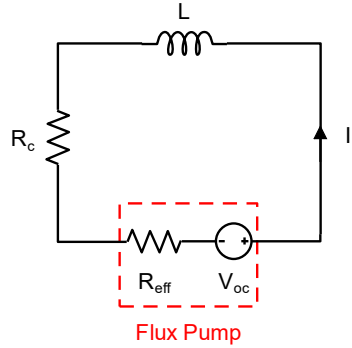}}
\caption {Equivalent electrical circuit model of an HTS dynamo connected to a coil via soldered joints.} 
\label{fig.circuit}
\end{figure}

For many configurations \cite{matairaR2020PRA,matairaR2020IES}, the $I-V$ curve of the flux pump is linear, and hence it can be modeled as a DC voltage source in series with an effective resistance, $R_{\rm eff}$, as shown in Fig. \ref{fig.circuit}. The value of the voltage source is equal to the DC open-circuit voltage of the dynamo $V_{oc}$.
During the operation, a coil with inductance $L$ is connected via the circuit resistance $R_c$ (resistance of soldered joints) to the dynamo. Therefore, the coil can be treated as an independent $LR$ circuit, which is charged by the voltage source. 

The current in the electrical circuit of Fig. \ref{fig.circuit} can be obtained using the following equation obtained by solving the governing equation of the circuit \cite{ZhenanJ2014APL}
 
\begin{eqnarray}
\label{equ.circuit}
i(t)=I_{sat}\left[1-e^{t/\tau}\right],
\end{eqnarray}
 
where $I_{sat}={V_{oc}}/({R_{c}+R_{\rm{eff}}})$ is the saturation current, which is the maximum value of current that the flux pump can deliver and $\tau=L/(R_{c}+R_{\rm{eff}})$ is the time constant of the circuit, determining the charging rate of the flux pump. The flux pump reaches 99.3\,$\%$  saturation of the final value, $I_{sat}$, after $5\tau$. 


\section{Results} \label{s.results}

The results of calculations using the MEMEP and SEG-H methods along with analytical results are presented and discussed in this section. 

For both methods, we have used 60 mesh elements across the width and one mesh element along the thickness of the tape. We have limited the number of mesh elements as much as possible up to a point that does not affect the accuracy of the results. The reason was to reduce the calculation time in order to be able to calculate up to millions of time steps in a reasonable time. 

\subsection{\textit{I-V} Curve}

To obtain the \textit{I-V} curve of the HTS dynamo, we impose different transport DC currents in the dynamo and calculate the corresponding DC voltage values.    
Using the \textit{I-V} curve, the HTS flux pump can be described as a current-controlled voltage source, as shown in Fig. (\ref{fig.circuit}). 

For our calculations, the transport current varies over the range $[0,I_{sc}]$, which outputs a DC voltage in the range of $[V_{oc},0]$. $I_{sc}$ is the transport current when the DC value of the output voltage is zero, i.e. short-circuit current and $V_{oc}$ is the output voltage when the transport current is zero, i.e. the open-circuit voltage. As mentioned before, we assumed an ideal superconducting coil. This indicates that in here, unlike experimental work where maximum current can be restricted by $I_c$ of the HTS coil, the maximum applied transport current here is $I_{sc}$, where the flux pump DC voltage becomes negligible. 

\begin{figure}[tbp!]
\centering
{\includegraphics[trim=0 0 0 0,clip,width=6.3 cm]{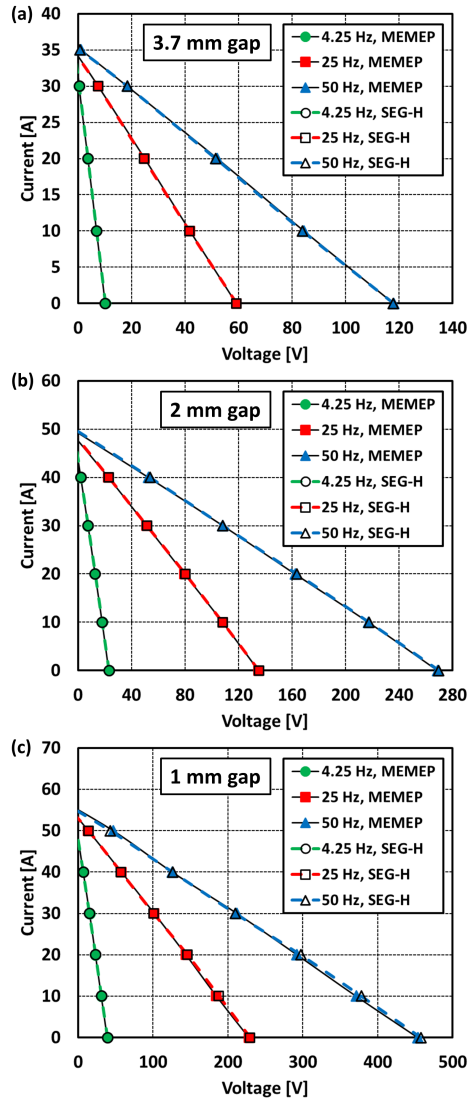}}
\caption {\textit{I-V} curves of the modeled HTS dynamo calculated by the MEMEP and SEG-H methods for three frequencies of 4.25, 25 and 50\,Hz and for airgaps of (a) 3.7\,mm, (b) 2\,mm, (c) 1\,mm.} 
\label{fig.I-V_curve}
\end{figure}

Fig. \ref{fig.I-V_curve} shows the \textit{I-V} curves of the modeled HTS dynamo calculated by the two numerical methods for three frequencies of 4.25, 25 and 50 Hz and three airgaps of 1, 2 and 3.7 mm. The results verify the fact already shown by experiments that $R_{\rm{eff}}$ for a fixed rotation frequency and superconducting operating regime has a constant value and is not a function of the current, which depends on the characteristic of the flux pump. In addition, the open-circuit voltage $V_{oc}$ and the slope of the \textit{I-V} curves, which describe the effective resistances of the dynamo, increase directly proportional to the frequency. The calculated effective resistance are summarized in Table \ref{Table.Reff}. The maximum percentage differences between the effective resistances of the models are only 0.6\,$\%$, 0.55\,$\%$ and 1.8\,$\%$ for the frequencies of 4.25 Hz, 25 Hz and 50 Hz, respectively, thus increasing with frequency.\;This difference increases slightly also by decreasing airgap. The values in the Table \ref{Table.Reff} show that regardless of the airgap value, the two compared methods have excellent agreement at low frequencies and with increasing the frequency, this agreement deteriorates slightly. 

\begin{table}[tbp]
\centering

\caption{Effective resistances (in $\mu\Omega$) calculated from the slope of the \textit{I-V} curves in Fig. \ref{fig.I-V_curve}}
\resizebox{9 cm}{!}{
\begin{tabular}{|c|c||c|c|c|} 
\toprule
\label{Table.Reff}
Method & \multicolumn{1}{c|}{\diagbox{Airgap}{Frequency}} & 4.25\,Hz & 25\,Hz & 50\,Hz  \\ 
\hhline{|==:t:===|}
MEMEP  & \multirow{2}{*}{1\,mm}                            & 0.803       & 4.289     & 8.138      \\ 
\cline{3-5}
SEG-H    &                                                  & 0.808       & 4.313     & 8.291      \\ 
\hhline{|--|:===|}
MEMEP  & \multirow{2}{*}{2\,mm}                            & 0.528       & 2.808     & 5.409      \\ 
\cline{3-5}
SEG-H    &                                                  & 0.528       & 2.807     & 5.34      \\ 
\hhline{|--|:===|}
MEMEP  & \multirow{2}{*}{3.7\,mm}                          & 0.324       & 1.723     & 3.336      \\ 
\cline{3-5}
SEG-H    &                                                  & 0.324       & 1.724    & 3.347      \\
\bottomrule
\end{tabular}
}
\end{table}

\subsection{Dynamic Charging of the HTS Dynamo}

In this section, we estimate and discuss the instantaneous output voltage components, the dynamic charging current curve of the coil and the resultant ripple AC loss. 
 
\subsubsection{Instantaneous voltage}
\label{Voltage-Dynamic charging}

\begin{figure}[tbp]
\centering
{\includegraphics[trim=0 0 0 0,clip,width=9 cm]{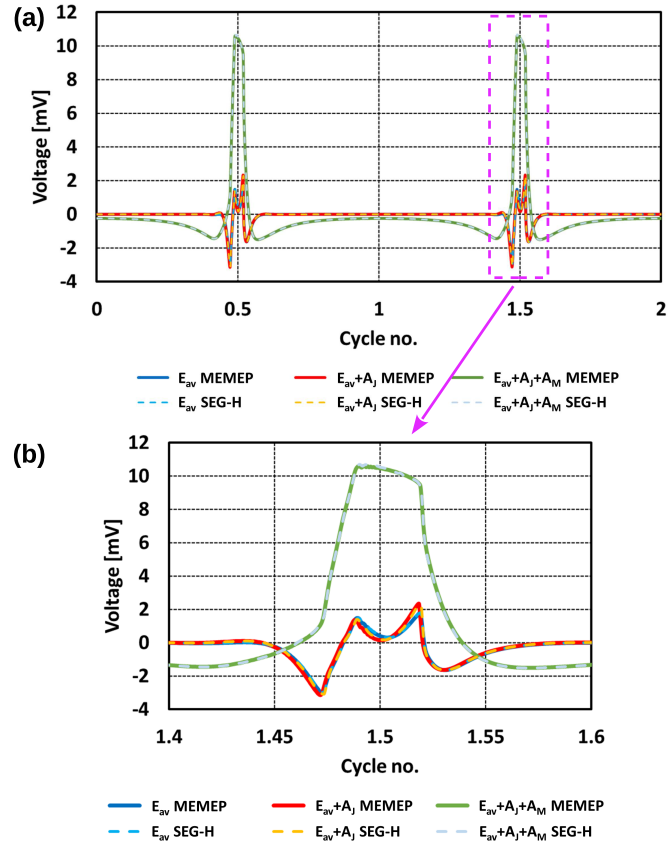}}
\caption {(a) Comparison of $l.E_{av}$ ($l\,\cdot$\,Equation (\ref{Eq.Eav})\,), $l \cdot [E_{av}+\partial_{t}A_{J,av}]$ ($l\,\cdot$\,[Equation (\ref{Eq.Eav})\,+\, $\partial_{t}A_{J,av}$]) and the total output voltage $V(t)$ (Equation (\ref{V_instan})\,) of the two studied methods for the first two cycles of the case with a 3.7\,mm airgap and a frequency of 25\,Hz (b) The magnified version of Fig. \ref{fig.zoom}\,(a) for the time period between 1.4 and 1.6 cycles.} 
\label{fig.zoom}
\end{figure}

Fig.\;\ref{fig.zoom}\,(a) shows a comparison of the three voltage components for the two numerical models for the first two cycles of the case with a 3.7\,mm airgap and a frequency of 25\,Hz including the average electric field multiplied by the tape effective length $l.E_{av}$, $l.(E_{av}+\partial_{t}A_{J,av})$, and the total output voltage $V(t)$. Fig. \ref{fig.zoom}\,(b) is the magnified version of Fig. \ref{fig.zoom}\,(a) for the time period between 1.4 and 1.6 cycle number. It is clear that the two models have very good agreement.           

\subsubsection{Instantaneous current}
\label{Current-Dynamic charging}
  
\begin{figure}[tbp]
\centering
{\includegraphics[trim=0 0 0 0,clip,width=7 cm]{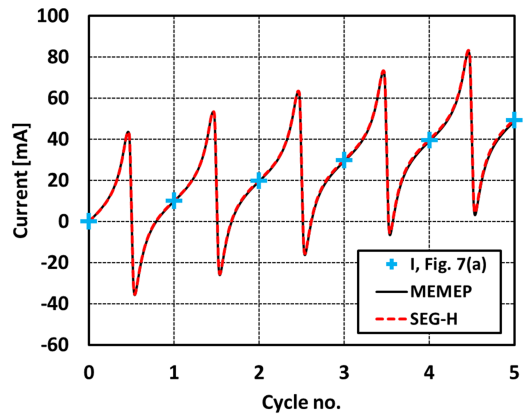}}
\caption {Dynamic charging current curve of the modeled coil over the first five cycles for the two studied models. The cross symbol refers to the extracted data points at the end of each cycle used for plotting Fig. \ref{fig.Charging_gap}. The case belongs to 3.7 mm airgap and 25 Hz frequency.} 
\label{fig.dynamic_charging}
\end{figure}

Fig. \ref{fig.dynamic_charging} shows the dynamic charging of the coil for the first five cycles of the case with a 3.7\,mm airgap and a frequency of 25\,Hz for the two studied models, which again have excellent agreement. The charging current curve contains ripples that resembles the ripples of cumulative total output voltage $V_{cumul}(t)$ of the HTS dynamo (Equation (\ref{V_cumul})\,).  
 
\subsubsection{Ripple AC loss}
\label{Ripple AC loss} 

\begin{figure}[tbp]
\centering
{\includegraphics[trim=0 0 0 0,clip,width=7 cm]{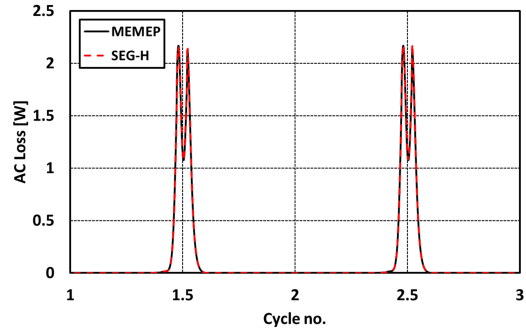}}
\caption {Ripple AC loss in the HTS tape of the modeled dynamo over the second and third cycles for the two studied models. The case belongs to the 3.7\,mm airgap and 25\,Hz frequency.} 
\label{fig.Ripple_AC_loss}
\end{figure}

The ripples of the charging curve shown in Fig. \ref{fig.dynamic_charging} generate AC loss in the HTS tape of the dynamo, which is shown in Fig. \ref{fig.Ripple_AC_loss}. We calculated the average AC loss in the first five charging cycles (ignoring the first transient cycle) for the MEMEP method as 135.4\,mW and for SEG-H method as 135.7\,mW. The average AC loss of the 5001$\textsuperscript{st}$ cycle, close to saturation of pumping, for the MEMEP method is 135.7\,mW and for SEG-H method is 135.9\,mW. This suggests that for a given frequency, the calculated ripple AC loss in the HTS dynamo is almost constant during the whole charging period of the coil.  

\subsection{Charging Behavior}
\begin{figure}[tbp!]
\centering
{\includegraphics[trim=0 0 0 0,clip,width=8.1 cm]{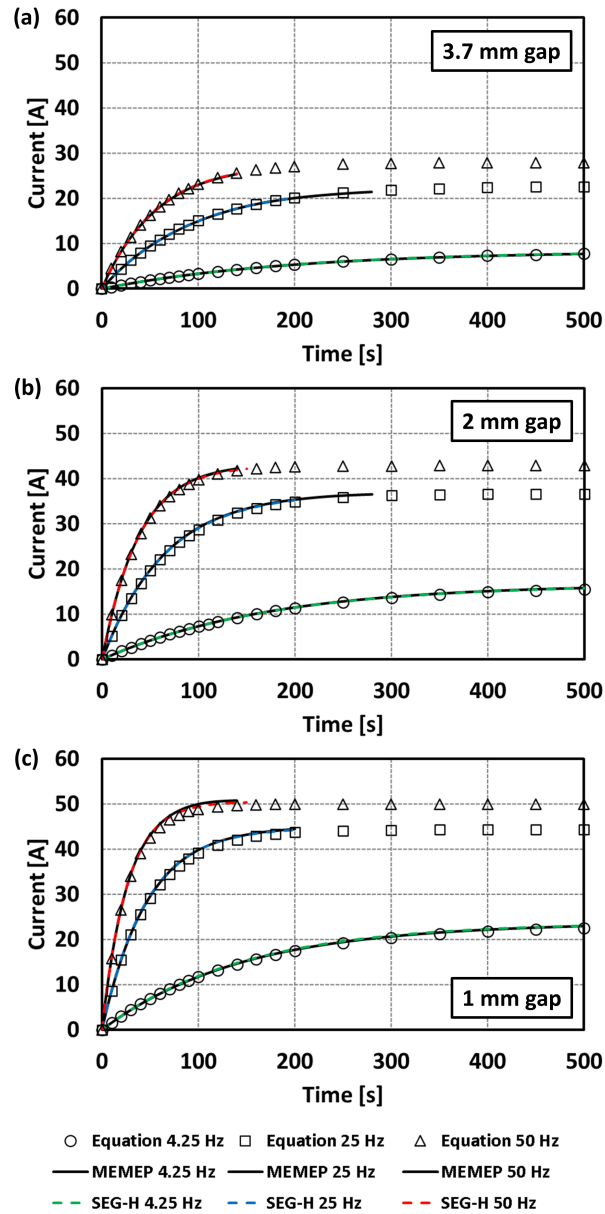}}
\caption {The calculated charging current curves of the coil for three frequencies of 4.25, 25 and 50\,Hz for airgaps of (a) 3.7\,mm, (b) 2\,mm, (c) 1\,mm. } 
\label{fig.Charging_gap}
\end{figure}

Fig. \ref{fig.Charging_gap} shows the calculated transport currents as a function of time for airgaps of (a) 3.7\,mm, (b) 2\,mm and (c) 1\,mm with rotation frequencies of 4.25, 25 and 50\,Hz, comparing the results of the two numerical methods with the analytical ones. To obtain a smooth charging curve in Fig. \ref{fig.Charging_gap} and to compare adequately with the analytical results, we extracted only the last data point of each cycle, where the magnet has returned to its starting position. 


Different number of cycles were required for each frequency to reach the saturation current $I_{sat}$, which is a function of the time constant $\tau$.\;For this reason, we used 2125, 5000 and 7000 calculated cycles for the frequencies of 4.25\,Hz, 25\,Hz and 50\,Hz, respectively, which correspond to durations of 500\,s, 200\,s and 140\,s. For the analytical calculations of the charging current, as given by the Equation (\ref{equ.circuit}), we have used the information in Table \ref{table_parameters} for the values of $R_c$ and $L$, the average values of $R_{\rm{eff}}$ obtained by the two numerical methods in Table \ref{Table.Reff} and also data obtained from Fig. \ref{fig.I-V_curve} for the values of $V_{oc}$.  

Fig. \ref{fig.Charging_gap} shows that for a given frequency, the coil current saturates at a higher value and faster as the airgap decreases. The shorter saturation time is due to the increase in $R_{\rm eff}$, since the time constant $\tau=L/(R_{\rm{eff}}+R_c)$ decreases with $R_{\rm eff}$. Therefore, $\tau$  increases with decreasing airgap, leading to faster charging of the coil. In the case of the saturation current, as shown in Equation (\ref{equ.circuit}), for $t\gg\tau$, $I_{sat}=V_{oc}/(R_{\rm{eff}}+R_c)$. As a consequence, with decreasing airgap, both $V_{oc}$ (from Fig. \ref{fig.I-V_curve}) and $R_{\rm{eff}}$ increases, but $V_{oc}$ increases comparatively more, leading to a resulting increase in $I_{sat}$. 

For a given airgap, the coil current saturates faster with a higher value of $I_{sat}$ as the frequency increases. $R_{\rm{eff}}$ increases linearly with frequency, thus the time constant $\tau$ decreases with increasing $R_{\rm{eff}}$. We also observe that $I_{sat}$ increases with increasing frequency, which suggests that $V_{oc}$ increases faster than $R_{\rm{eff}}$ as the frequency increases.

By considering the last calculated data point of each current curve in Fig. \ref{fig.Charging_gap} as the criterion, we investigated the agreement between the models and the analytical method. The maximum percentage difference between MEMEP and SEG-H method at frequencies of 4.25\,Hz, 25\,Hz and 50\,Hz was only 0.7\,$\%$, 0.4\,$\%$ and 0.96\,$\%$, respectively. The maximum percentage difference between SEG-H and the analytical method at frequencies of 4.25\,Hz, 25\,Hz and 50\,Hz was 1.7\,$\%$, 1.2\,$\%$ and 1.6\,$\%$ and between MEMEP and the analytical method were 2.1\,$\%$, 1.65\,$\%$ and 2.13\,$\%$, respectively. Therefore, we conclude that there is an good agreement between both numerical models and with the analytical model. In addition, the charging current curves have good qualitative agreement with the curves obtained by experiment presented in \cite{ZhenanJ2014APL, ZhenanJ2015SUST}.   

While the voltage signals in Fig. \ref{fig.zoom} look quite different, the DC component is the same and this leads to our ability to estimate the charging behavior per cycle based on average values, leading to the derivation of the \textit{I-V} curves in Fig. \ref{fig.I-V_curve} and the ability to use the analytical equation presented earlier.     

\subsection{Discussion}

One may question why such complications are needed to analyze the charging process of an HTS dynamo, when it can be modeled simply by the analytical considerations presented in Section \ref{Analytical method}. For this purpose, only the \textit{I-V} curve of the flux pump (including the effective resistance of the flux pump, which is still obtained from numerical models) and the characteristics of the coil would be needed. However, the analytical model cannot capture the dynamic behavior of the flux pump and the output voltage. This can be important and is only possible when the HTS dynamo is coupled with the coil circuit, as it has been performed in these models. In addition, our models have the potential to be coupled with other multiphysics analyses, as well as with a MEMEP or finite-element model of the superconducting coil in more detail. Furthermore, the models allow investigations on HTS dynamos with nonlinear \textit{I-V} characteristics and their influence on the dynamics of the charging process \cite{matairaR2020IES}, which is not possible to study via the analytical method.  

Crucially, the current ripple shape in each cycle reflects the dynamic operation of the HTS dynamo while the magnet transits past the tape, resulting in the two peaks of the current. Since these ripples can potentially cause AC loss and subsequent thermal loss in the coil, accurate calculations of these ripples are important. 

\section{Conclusion}

In this work, two novel numerical methods for simulating the charging process of a coil by an HTS dynamo were presented: The Minimum Electromagnetic Entropy Production (MEMEP) and the segregated $\textbf{H}$-formulation (SEG-H). They were compared together and with an analytical method through nine different cases including three airgaps of 1, 2 and 3.7\,mm and three frequencies of 4.25, 25 and 50\,Hz.~Firstly, the \textit{I-V} curves of the modeled HTS dynamo and thus the effective resistances were calculated and compared together.~The maximum percentage difference between the effective resistances of the two models was less than 2\,$\%$, which occurred at the highest frequency and the smallest airgap. Then, the instantaneous voltage components and the dynamic charging current curve of the coil for the two models were calculated and compared together, which showed excellent quantitative and qualitative agreement. It was found that the current charging curve contains ripples within each cycle, which cannot be captured via the analytical method.~Such ripples cause ripple AC loss in the HTS dynamo, which was shown to be almost constant during the whole charging process. Afterwards, the charging process of a coil by the dynamo was investigated and there was again excellent quantitative and qualitative agreement between the two models and the analytical method. In all, the two presented numerical methods showed promising performance to describe the charging process of an HTS dynamo over thousands of cycles, as well as capturing the current ripple within a cycle. In addition, the flexibility of the numerical modeling frameworks presented here have the potential to be coupled with other multiphysics analyses, as well as with a MEMEP or finite-element model of an HTS coil. Besides, the models are capable of studying HTS dynamos with nonlinear \textit{I-V} characteristics and their influence on dynamics of the charging process.

\section*{Acknowledgment}
M D Ainslie would like to acknowledge financial support from an Engineering and Physical Sciences Research Council (EPSRC) Early Career Fellowship, EP/P020313/1. Additional data related to this publication are available at the University of Cambridge data repository (\url{https://doi.org/10.17863/CAM.66975}). This work also received the financial support of the Slovak grant agencies APVV with contract number APVV-19-0536 and VEGA with contract number 2/0097/18. 

\section*{References}

\bibliographystyle{unsrt}

\bibliography{local.bib}

\end{document}